\documentclass[twocolumn,prl,aps,superscriptaddress,showpacs]{revtex4-1}

 \usepackage{mathptmx}
\usepackage{dcolumn}
\usepackage{amsmath,amssymb,amsfonts}
\usepackage{dsfont}
\usepackage{bm}
\usepackage{color}
\usepackage{latexsym}
\usepackage{epstopdf}
\usepackage[english]{babel}
\usepackage{enumerate}

\usepackage[pdftex]{graphicx}

\usepackage{natbib}
\usepackage{epstopdf}
\usepackage{appendix}

\definecolor{mygrey}{gray}{0.35}
\definecolor{myblue}{rgb}{0.2,0.2,0.8}
\definecolor{myzard}{cmyk}{0,0,0.05,0}
\definecolor{mywhite}{rgb}{1,1,1}
\definecolor{mywhite}{rgb}{1,1,1}
\definecolor{myred}{rgb}{1,0.,0.3}
\usepackage[colorlinks=true,citecolor=myblue,linkcolor=myred]{hyperref}

\newcommand{\bra}[1]{\left\langle #1\right|}
\newcommand{\ket}[1]{\left| #1\right\rangle}

\newcommand\nn{\mathbf{n}}
\newcommand\mm{\mathbf{m}}
\newcommand\kk{\mathbf{k}}
\newcommand\qq{\mathbf{q}}

\newcommand\intt{\mathrm{int}}

\begin{document}

\title{Quantum Emitters in Two-dimensional Structured Reservoirs in the Non-Perturbative Regime}
 \author{A. Gonz\'{a}lez-Tudela}
 \email{alejandro.gonzalez-tudela@mpq.mpg.de}
 \affiliation{Max-Planck-Institut f\"{u}r Quantenoptik Hans-Kopfermann-Str. 1. 85748 Garching, Germany }
\author{J. I. Cirac}
 \affiliation{Max-Planck-Institut f\"{u}r Quantenoptik Hans-Kopfermann-Str. 1. 85748 Garching, Germany }

\begin{abstract}
We show that the coupling of quantum emitters to a two-dimensional reservoir with a simple band structure gives rise to exotic quantum dynamics with no analogue in other scenarios and which can not be captured by standard perturbative treatments. In particular, for a single quantum emitter with its transition frequency in the middle of the band we predict an exponential relaxation at a rate different from that predicted by the Fermi's Golden rule, followed by overdamped oscillations and slow relaxation decay dynamics. This is accompanied by directional emission into the reservoir. This directionality leads to a modification of the emission rate for few emitters and even perfect subradiance, i.e., suppression of spontaneous emission, for four quantum emitters.
\end{abstract}

\maketitle

The interaction of quantum emitters (QEs) with propagating bosonic particles, e.g., photons, lies at the core of Quantum Optics \cite{cohenbook92a}. This interaction leads, for example, to collective interactions between QEs \cite{lehmberg70a,lehmberg70b} which can be harnessed for both quantum information and simulation applications. New avenues in the integration of QEs with nanophotonic structures \cite{vetsch10a,huck11a,hausmann12a,laucht12a,thompson13a,goban13a,lodahl15a,goban15a,sipahigi16a,bermudez15a,beguin14a} provide us with systems in which the QEs interact with low dimensional bosonic modes, with complicated energy dispersions in the case of engineered dielectrics \cite{hausmann12a,laucht12a,thompson13a,goban13a,lodahl15a,goban15a,sipahigi16a}. Despite originally the main motivation of such implementations was to exploit the small sizes to enhance light-matter interactions, it was soon realized that intriguing phenomena arise because of the reduced 
dimensionality. One particular aspect is the possibility of realizing chiral emission~\cite{mitsch14a,sollner15a,lodahl16a}, which can display very uncommon features~\cite{ramos14a,pichler17a}. Another one is the possibility of exploiting the phenomena of sub and superradiance~\cite{goban15a,corzo16a,sorensen16a,sipahigi16a,solano17a}, e.g., to enhance the coupling to the emitter~\cite{chang12a}, to generate QE entanglement~\cite{gonzaleztudela13a,ramos14a}, to produce non-classical light~\cite{gonzaleztudela15a,gonzaleztudela16a}, or even to perform quantum computation~\cite{paulisch16a}.

The dynamics of QEs in 1D reservoirs is relatively simple, specially when their transition frequency, $\omega_e$, lies within a band. Perturbative treatments predict that a single QE initially excited decays at a rate, $\Gamma$, given by the Fermi's Golden Rule (FGR), i.e. proportional to the density of states of the bath at $\omega_e$. The emission mostly occurs in the bath modes that are resonant with that frequency. Typically, there are two such modes of associated momentum $\pm k_e$, leading to a symmetric left/right emission. When two (or more) QEs are present, the existence of only two such modes leads to super/subradiant states, where the emission is enhanced or suppressed by interference. In higher dimensions for structureless baths, a single QE will also decay at a rate given by the FGR. However, the emission takes place in different directions as there are many resonant modes in the bath. For two QEs, the interference in the emission cannot occur in all those modes at the same time \footnote{Unless their separation is smaller than the 
associated wavelength~\cite{dicke54a}, something we are not considering here} and thus, the phenomena of sub and superradiance are generically absent. 

In this manuscript we use non-perturbative methods to analyze the dynamics of QEs interacting with a simple two-dimensional (2D) structured reservoir. By structured we mean with a periodic structure giving rise to a dispersion relation containing frequency bands. In particular, we contrast our results with the predictions of perturbative treatments based on a markovian master equation approach. First, we show how a single excited QE with its energy tuned in the middle of the band shows an exponential decay at a rate different from that predicted by FGR. Moreover, at longer times, this exponential relaxation is followed by an oscillation and subexponential dynamics. These QE dynamics are followed by a directional emission into the bath in two orthogonal quasi-1D directions, as also predicted for classical sources of light~\cite{mekis99a} and sound~\cite{langley96a}. As a consequence, when several QEs are coupled to the bath, this directional emission induces anisotropic collective dissipation. For two QEs we observe a modification of the spontaneous emission rate as a consequence of this directional emission when the QE lie on a line at $45$ degrees. A related behavior has been predicted in \cite{galve17a} using perturbative master equations. We find that a total suppression of the emission is not possible for two QEs, and explain this fact in terms of a partial interference effect. In contrast, we show how to design a perfect subradiant state with four QEs which survives even in the non-perturbative regime, since in that case the interference can be fully destructive,

\begin{figure}
\centering
\includegraphics[width=0.9\linewidth]{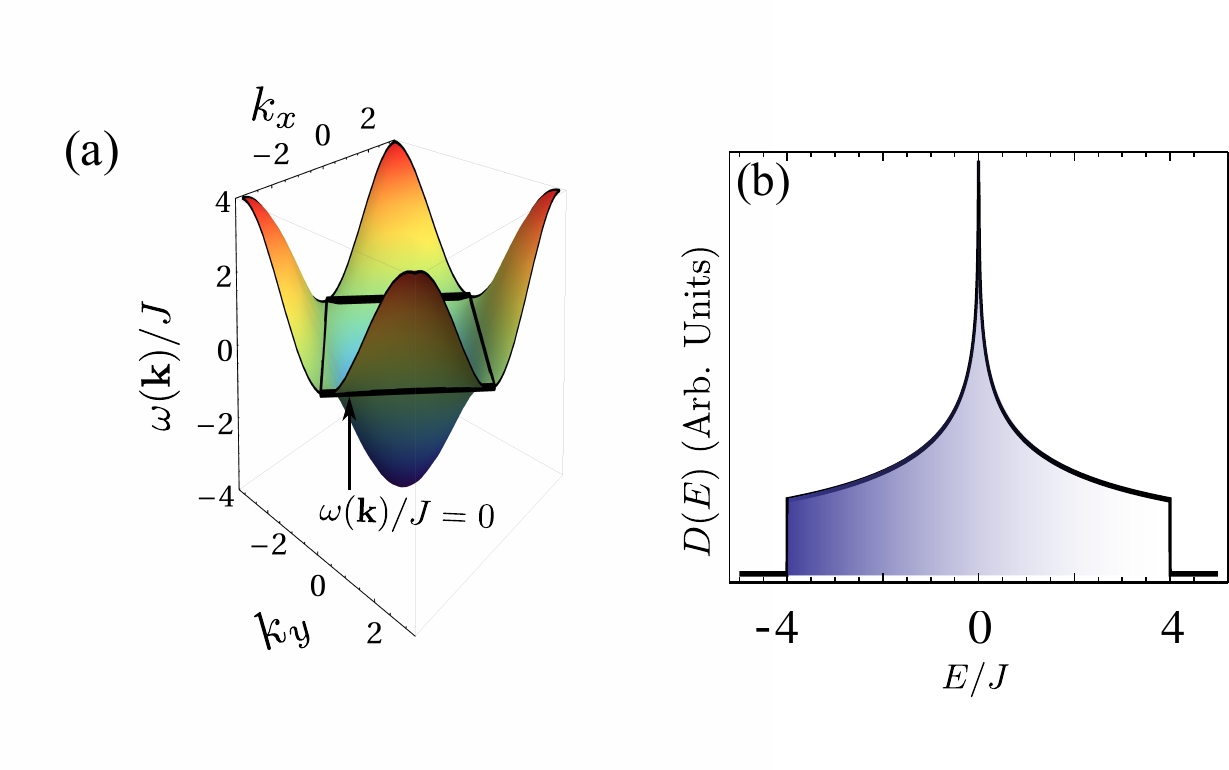}
\caption{ (a) [(b)] Energy dispersion $\omega(\kk)/J$ [and density of states $D(E)$] for the square lattice tight-binding model with a black line highlighting the $\kk$ points satisfying $\omega(\kk)=0$.}
\label{fig1}
\end{figure}

We assume a 2D bath with a square-like symmetry described by $N\times N$ bosonic modes with energy $\omega_a$ and with nearest neighbour coupling $J$. The Hamiltonian is given by (using $\hbar=1$) $H_B=-J\sum_{\langle \nn,\mm \rangle} \left(a_\mm^\dagger a_\nn+\mathrm{h.c.}\right)$, where $\nn=(n_x,n_y)$ is a vector indicating the bosonic mode position within the lattice. We have used a rotating frame at a frequency $\omega_a$, such that the zero energy corresponds to the center of the band structure (see below). Despite the simplicity of the model, we expect it to describe more complex materials such as photonic crystals, in the same way that the 1D tight-binding model does for structured waveguide QED. The bath Hamiltonian can be diagonalized in $\kk$ space by introducing periodic boundary conditions and the 
operators $a_\kk=\frac{1}{N}\sum_\nn e^{-i \kk \nn } a_\nn$, where $\kk=(k_x,k_y)$ 
and $k_x,k_y=\frac{2\pi}{N}(-\frac{N}{2},\dots,\frac{N}{2}-1)$, 
such that $H_B=\sum_\kk \omega(\kk) a_\kk^\dagger a_\kk$ with $\omega(\kk)=-2 J \left[\cos(k_x)+\cos(k_y)\right]$. In Fig.~\ref{fig1}, we plot the energy dispersion of the band $\omega(\kk)$ together with the associated density of states in the limit $N\rightarrow\infty$, $ D(E)=\frac{1}{(2\pi)^2}\iint d\kk \delta[E-\omega(\kk)]$. The band extends from $[-4J,4J]$. At the band edges, the $D(E)$ is nearly constant as predicted for isotropic dispersions~\cite{gonzaleztudela15c}. At the middle of the band it displays a divergence associated to the saddle point appearing at the energy dispersion $\omega(\kk)$, as it also happens for real materials~\cite{mekis99a}. As we show below, this has important consequences in the dynamics beyond purely enhancing the emission. We also consider one (or several) QEs described as two-level systems $\{\ket{g}_j,\ket{e}_j\}$, with transition frequency, $\omega_e$,
whose Hamiltonian reads: $H_S=\Delta \sum_j  \sigma^j_{ee}$. We use the notation $\sigma^j_{\alpha\beta}=\ket{\alpha}_j\bra{\beta}$ for the spin operators and $\Delta=\omega_e-\omega_a$ represents the detuning with respect to the middle of the band. Finally, we assume a local coupling of the QEs to the bath modes, described by $H_\mathrm{int}=g\sum_j \left( a_{\nn_j}\sigma_{eg}^j+\mathrm{h.c.}\right)$. We assumed to be in a parameter regime where the QEs are coupled to a single band of the bosonic reservoir.

Along this manuscript we consider the QE(s) to be initially excited in certain QE state $\ket{\Phi_0}_S$, whereas the bath starts initially empty, i.e., $\ket{\mathrm{vac}}_B=\ket{0}^{\otimes N^2}$. Then, we let the system free to evolve under the total Hamiltonian $H=H_S+H_B+H_\intt$, and study the QEs relaxation. The situation when the QE energies lie outside the band, $|\Delta\pm 4J|\gg g$, where the dynamics are dominated by the presence of a bound state, has been explored extensively in other works (see, e.g.,~\cite{bykov75a,john90a,kurizki90a}). We focus here instead on the situation when the QE energies lie within the band, $\Delta\in [-4J,4J]$, with emphasis on what happens in the middle of the band. In our illustrations, we use relatively large values of $g$ in order to emphasize the features in the figures; however, the conclusions can be extended to smaller $g$'s that is the common situation in the optical regime. Nevertheless, we 
make a complete discussion about the whole range of parameters and give more details about the calculation in the accompanying paper~\cite{gonzaleztudela17b}.

We first consider that a single QE is coupled to the bath, such that we drop the index $j$ in the QE operators, and assume $\ket{\Phi_0}_S=\ket{e}$. As the initial state contains a single excitation, the state at any time, $t$, can be written as follows:
\begin{equation}
  \ket{\Psi(t)} = \left[C_e(t)\sigma_{eg} + \sum_\nn C_\nn(t) a_\nn^\dagger\right]\ket{g}\otimes\ket{\mathrm{vac}}_B, 
\end{equation}

\begin{figure}
\centering
\includegraphics[width=0.7\linewidth]{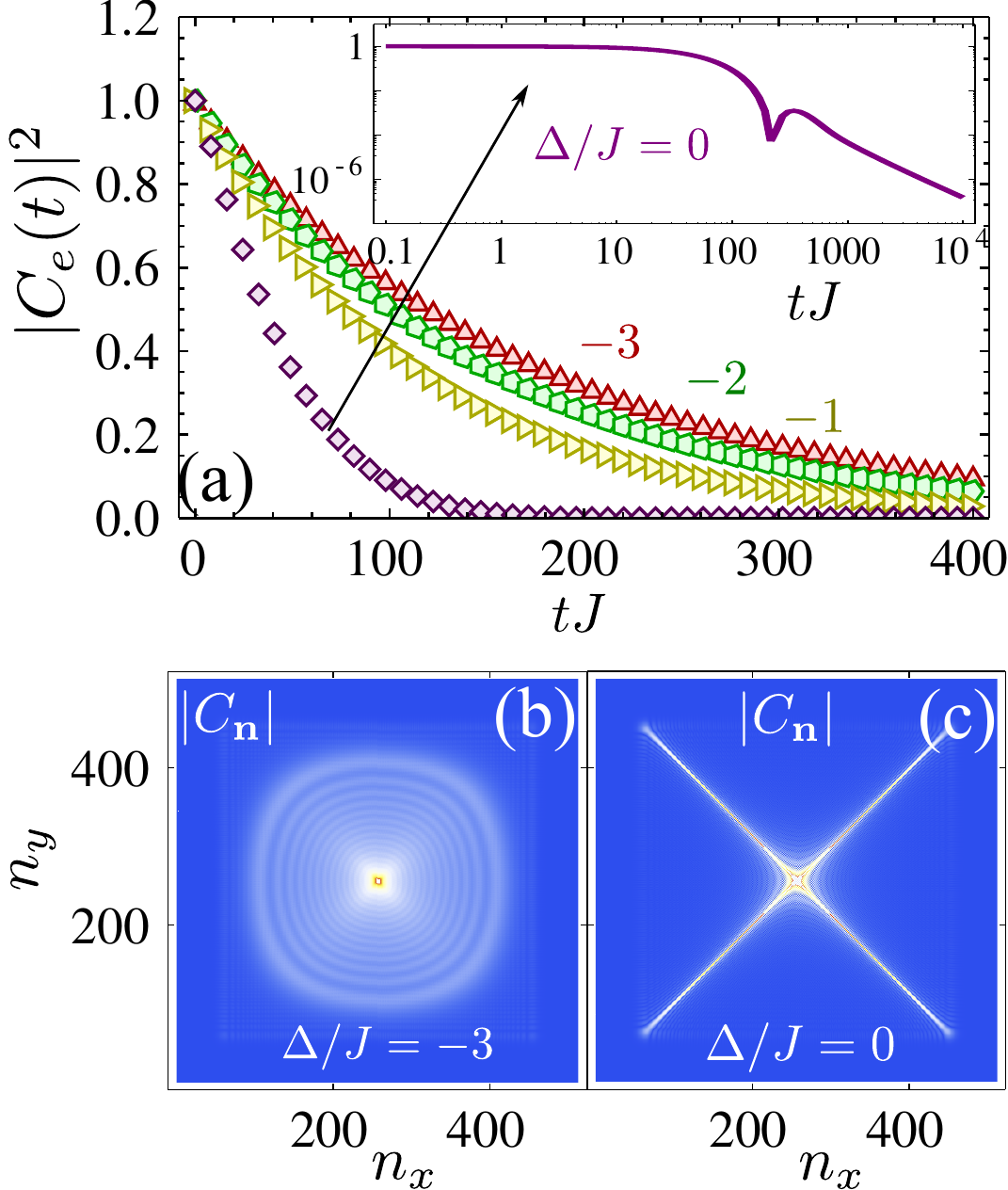}
\caption{(a) Excited state population $|C_e(t)|^2$ for a single QE for $g/J=0.1$ and different QE energies as depicted in the legend. Inset:  Excited state population $|C_e(t)|^2$ in logarithmic scale for $\Delta=0$ to visualize the non-perturbative dynamics. (b-c) Probability amplitude of the bath modes, $|C_\nn|$, for positions $\nn=(n_x,n_y)$ at a time $tJ =100$ for $\Delta/J=-3$ and $0$ respectively. }
\label{fig2}
\end{figure}

In Fig.~\ref{fig2} we show the results of the numerical integration of the dynamics of the QEs for several detunings for $\Delta/J=-3,-2,-1,0$ and $g=0.1J$, which we complement with the state of the bath excitations, i.e., $|C_{\nn}|$, at $tJ=100$ for $\Delta=-3J,0$. From Fig.~\ref{fig2}(a), it seems that the decay of the QE is basically exponential, with an enhanced decay rate as its energy is tuned closer to the center of the band. Naively, this is what one expects from perturbative approaches, which predict $|C_e(t)|^2 \approx e^{-\Gamma_e(\Delta)t}$, with a decay rate given by FGR, $\Gamma_e(\Delta)=2\pi g^2 D(\Delta)$, where $D(\Delta)$ is the density of modes evaluated at the QE energy $\Delta$. For $\Delta/J=-3,-2,-1$ such prediction works well; however, for $\Delta=0$ an straightforward application of FGR predicts an infinite decay rate that we do not observe in Fig.~\ref{fig2}(a). Moreover, when plotting the population dynamics in 
logarithmic scale (inset) we observe that the relaxation is actually non-monotonic 
showing both an oscillation and a subexponential decay for long times. 

To gain analytical insight into this exotic relaxation, we apply standard techniques~\cite{cohenbook92a} to rewrite the probability amplitudes in terms of their Fourier transform $C_{e,\kk}(t)=\frac{i}{2\pi}\int_{-\infty}^{\infty} d E G_{e,\kk}(E+i0^+) e^{-i E t}$, where:
\begin{align}
\label{eq:singleQEdyn}
 G_{e}(z)=\frac{1}{z-\Delta-\Sigma_e(z)}\,,
 G_\kk(z)=g\frac{G_e(z)}{z-\omega(\kk)}\,,
\end{align}
and $\Sigma_e(z)=\frac{g^2}{N^2}\sum_\kk \frac{1}{z-\omega(\kk)}$ is the so-called QE \emph{self-energy} that captures the effect of the coupling to the bath on the QE dynamics. From now on we focus on what happens for energies, $E$, around the middle of the band ($|E|\ll J$) where the self-energy can be expanded as~\cite{morita71a,economoubook83a,gonzaleztudela17b}
\begin{equation}
  \Sigma_{e}(E+i0^+)\approx \frac{g^2}{4 J}\left[\mathrm{sgn}(E)- \frac{ 2 i}{\pi}\log\left(\frac{|E|}{16J}\right) \right].\,
\end{equation}

In this expression we observe that around $E=0$ the real part, that we denote as $\delta\omega_e(E)$, has a discontinuous jump, whereas the imaginary one, denoted as $\Gamma_e(E)$, has a logarithmic divergence. For the latter, we use the same notation as the FGR decay rate because they are connected. More concretely, the standard perturbative approaches, such as the Markov approximation~\cite{gardiner_book00a}, assume the self-energy to smoothly vary around $\Delta$ and replace $\Sigma_e(E+i0^+)\approx \Sigma_e(\Delta+i 0^+)$, recovering the exponential relaxation of $|C_e(t)|^2$ with decay rate, $\Gamma_e(\Delta)$, as given by the FGR.

The divergence appearing in the middle of the band, however, forces us to go beyond the perturbative treatment to unravel the results. In the standard quantum optical scenario \cite{cohenbook92a} one calculates the exact Fourier transform of $C_\alpha(t)$ by closing the contour in the lower half complex plane $\mathrm{Im}(E)<0$, taking detours at the band edges of $\omega(\kk)$ because of the presence of branch cuts in the self-energy. The poles in the real axis describe bound states, and give rise to fractional decay~\cite{john94a}. The  complex ones lead to an exponential decay. Finally, branch cuts associated to the band-edges give rise to power-law decays, which are, however, typically hidden by the fractional decay induced by bound states~\cite{john94a}. In the case studied here, an additional branch cut appears in the (negative) imaginary axis, associated to the divergence of the $D(E)$ in the middle of the band. This forces us to take an extra detour in the integration contour which has two visible 
consequences in the inset of Fig.~\ref{fig2}(a): i) the slow relaxation dynamics at long times scaling as $O[\left(t\log(16Jt)^2\right)^{-2}]$; ii) there exists a regime of $\Delta\in [-\frac{g^2}{2J},\frac{g^2}{2J}]$, in which two unstable poles appear in the analytical continuation of $G_\alpha(z)$. For $\Delta=0$, their imaginary part coincides and is given by~\cite{gonzaleztudela17b}
\begin{align}
\label{eq:Gammae}
 \bar{\Gamma}_e\approx \frac{g^2}{\pi J}\log\left(\frac{32\pi J^2}{g^2}\right)
\end{align}
obtained in the limit $\frac{32\pi J^2}{g^2}\gg 1$. Their real part is given by $\approx \pm \frac{g^2}{2J}$. This explains both the finite time scale observed in Fig.~\ref{fig2}(a), and the oscillation observed in the inset, which has a frequency proportional to the difference of the real part of the unstable poles ($\sim \frac{g^2}{J}$). Both behaviours emerge from the failure of perturbative treatments due to the divergent density of states. The oscillations can also be intuitively understood as part of the $\kk$-modes propagate very slowly allowing them to be reabsorbed by the QE. Remarkably, for short times one observes an exponential relaxation with a timescale given by Eq.~\ref{eq:Gammae} rather than the one expected from perturbative arguments that would be $O(g^2/J)$.

We also plot the emission into the bath in Figs.~\ref{fig2}(b-c). The bath population can also be obtained exactly in the long-time limit, as the probability amplitude $C_\kk(t)$ is still dominated by the pole contribution of $G_\kk(E+i0^+)$ of Eq.~\ref{eq:singleQEdyn} at $E=\omega(\kk)$, which yields:
\begin{equation}
 \label{eq:ck}
 \lim_{t\rightarrow\infty} C_\kk(t)=\frac{g e^{-i \omega(\kk) t}}{\omega(\kk)-\Delta-\Sigma_e(\omega(\kk))}\,,
\end{equation}

This expression tells us that the $\kk$-modes around $\omega(\kk)\approx \Delta$ for $g\ll J$ are the ones dominating the emission. This explains why when $\Delta$ is away from zero, the $\kk$ modes are isotropically populated, as $\omega(\kk)\approx f(|\kk|^2)$. For $\Delta\approx 0$, we have that the modes that dominate fulfill $ k_x\pm k_y=\pi$, as sketched in Fig.~\ref{fig1}(a). This explains why the emission is anisotropic in this case: the only resonant modes fulfill $k_x \pm k_y\approx \pm \pi$, and thus propagate along the diagonals. 

\begin{figure}
\centering
 \includegraphics[width=0.7\linewidth]{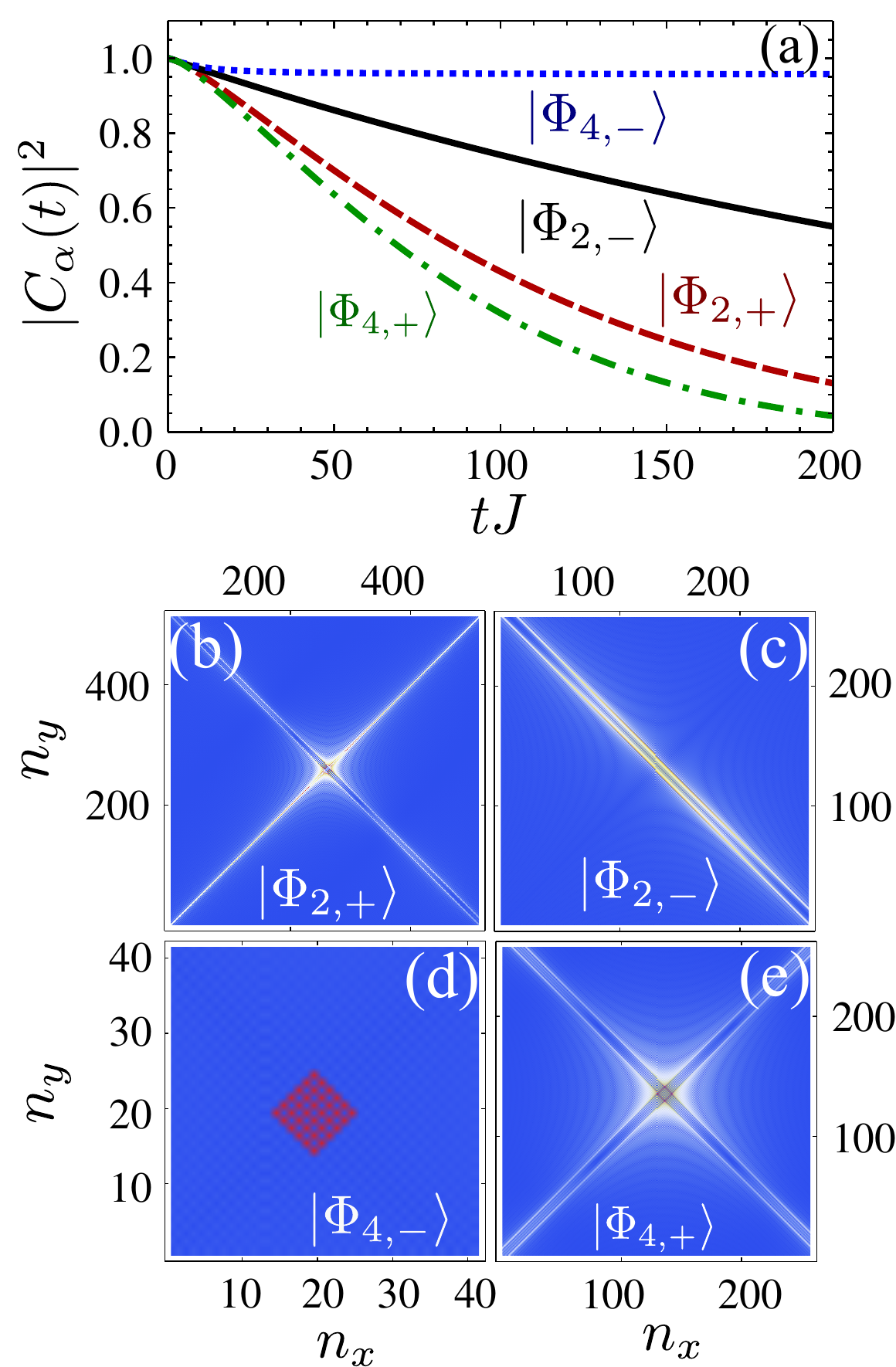}
\caption{(a) Population of states $\ket{\Phi_{2,\pm}}$ [$\ket{\Phi_{4,\pm}}$] for a coupling $g=0.05J$ for positions $\nn_{12}=(6,6)$ [and $(6,0),(0,6), (6,12), (12,6)$] respectively. (b-e) Bath probability amplitude at time $tJ=200$ for the initial states of panel (a) as depicted in the legend.}
\label{fig3}
\end{figure}

Now, we include several QEs in the discussion and take $\Delta=0$. We explore the interplay between the anisotropic emission and the relative position, $\nn_{12}$ of the QEs, to check up to which point super/subradiance phenomena survive within a non-perturbative picture. We first study the scenario with two QEs prepared in a (anti)symmetric superposition $\ket{\Phi_0}_S=\ket{\Phi_{2,\pm}}=\frac{1}{\sqrt{2}}\left(\sigma_{eg}^1\pm\sigma_{eg}^2\right)\ket{g}^{\otimes 2}$. Interestingly, when $\omega(\kk)=\omega(-\kk)$ the dynamics of the symmetric/antisymmetric state separate because they couple to orthogonal bath modes. Consequently, their relaxation dynamics is calculated analogously to the one of a single QE, but replacing: $\Sigma_e\rightarrow \Sigma_\pm=\Sigma_e\pm \Sigma_{12}$, where $\Sigma_{12}=\frac{g^2}{N^2}\sum_\kk \frac{e^{i\kk\cdot\nn_{12}}}{z-\omega(\kk)}$ is the collective QE interaction induced by the environment, which can be obtained recursively \cite{morita71a,economoubook83a,
gonzaleztudela17b}. Along this manuscript, we focus on a situation when the two QEs lie along a diagonal, $\nn_{12}=(n,n)$, as this is where most of the emission occurs [see Fig.~\ref{fig2}(c)], and may lead to modifications of collective decay~\cite{galve17a}. We find that that perturbative Markov approaches, i.e., $\Sigma_{12}(i0^+;\nn)/\Sigma_{e}(i0^+)=(-1)^n$, which point to the possibility of perfect super/subradiance. However, from our single QE study we know at $\Delta=0$ perturbative approaches may fail.

To go beyond perturbative treatments, we first numerically integrate the dynamics and show that if $n$ is even (odd), the states $\ket{\Phi_{\pm (\mp)}}$ have enhanced (suppressed) decay rates, as expected from the propagation phases of the bath modes satisfying $k_x\pm k_y=\pi$, which are the ones dominating the dynamics when $\Delta=0$. We show an example of such dynamics in Figs.~\ref{fig3}(a-b) for $\nn_{12}=(6,6)$, where we observe collective effects leading to enhancement/suppression of spontaneous emission. Notice, however, that neither the suppression nor the enhancement is perfect as it occurs in 1D systems. The reason behind that can be intuitively understood from the bath population that we plot in Figs.~\ref{fig3}(b-c). For the superradiant state, i.e., $\ket{\Phi_{2,+}}$, the emission occurs in 3 quasi-1D modes, one collective along the diagonal where the QEs are placed and two independent ones along the orthogonal directions. On the contrary, the subradiant state $\ket{\Phi_{2,-}}$ suppresses 
the emission in the diagonal where they are placed, while emitting into (two) quasi-1D modes along the orthogonal diagonals. These independent decay channels forbid finding a perfectly subradiant state with only two QEs. This enhancement/suppression can also be explained in terms of constructive/destructive interference of the bath modes emitted by the two QEs at $\omega(\kk)=0$, which are mainly given $k_x+k_y=\pm \pi$, such that their phases: $e^{i (k_x+k_y)n}\pm e^{i (k_x+k_y)n}=e^{\pm i\pi n}\pm e^{\pm i\pi n}$, add up constructively/destructively depending on the relative phase between QEs. Apart from that, we observe: i) retardation effects, as they also occur in 1D systems; ii) other non-Markovian effects introduced by the divergence of the density of states, which lead to a logarithmic correction of the super/subradiant decay with the distance~\cite{gonzaleztudela17b}. 

The intuition obtained with two QEs allows us to build perfect subradiant states with four QEs. Let us consider four QEs at positions $(0,2n)$, $(2n,0)$, $(2n,4n)$ and $(4n,2n)$ and assume they are in a state  $\ket{\Phi_{4,\pm}}=\frac{1}{2}\left(\sigma_{eg}^1\pm \sigma_{eg}^2+\sigma_{eg}^3\pm \sigma_{eg}^4\right)\ket{g}^{\otimes 4}$, The decay of $\ket{\Phi_{4,+}}$ is collectively enhanced with an emission in $8$ orthogonal directions as shown in Fig.~\ref{fig3}(d). Starting out in $\ket{\Phi_{4,-}}$, the emission is completely suppressed (up retardation effects) because perfect destructive interference occurs in the eight emission directions, trapping the light between the four QEs [see Fig.~\ref{fig3}(d)]. The analysis to show that the $\ket{\Phi_{4,-}}$ is perfectly subradiant relies on the fact that in this position configuration, the QE modes defined by $\left(\sigma_{eg}^1\pm\sigma_{eg}^2+\sigma_{eg}^3\pm \sigma_{eg}^4\right)$ couple to orthogonal bath modes, that allow us to consider their dynamics 
as 
those of a single QE, but with a modified self-energy. In particular, for 
the subradiant case the associated self-energy can be shown to be~\cite{gonzaleztudela17b}:
\begin{equation}
 \Sigma_{4,-}(z)=\frac{4g^2}{\pi^2}\iint_{0}^\pi d\qq \frac{\sin^2(2q_x n)\sin^2(2q_y n)}{z+4J \cos(q_x)\cos(q_y)}\,,\nonumber
\end{equation}
where we have used the rotated axis coordinates $k_{x,y}=q_x\pm q_y$. If $\Sigma_{4,-}(z)$ vanishes at $z=0$, this implies that a real bound state emerges within the band. This can be shown to be the case because the integrand at $z=0$ is separable in $q_{x,y}$, and each integrand satisfies: i) $I(q_{x,y})=-I(\frac{\pi}{2}-q_{x,y})$ and ii) its divergence appearing because of the zeros in the denominator for the $\qq$ modes satisfying $q_{x,y}=\pi/2$ is canceled by the one in the numerator. The last thing to show is that, indeed, its associated residue, connected to the steady-state population, is not zero. We can explicitly calculate it obtaining:
\begin{equation}
 C_{4,-}(\infty)=\frac{1}{1-\partial_z\Sigma_{4,-}(z)}\Big|_{z=0}=\frac{1}{1+\frac{g^2 n^2}{J^2}}\,.
 \end{equation}
which is therefore very close to $1$ as long as retardation effects are small $\frac{g^2 n^2}{J^2}\ll 1$. In any case, we obtain that independent of the distances there always remains some excitation within the QEs.

Finally, let us briefly comment how the Hamiltonian $H$ can be obtained on a platform beyond engineered dielectrics~\cite{hausmann12a,laucht12a,thompson13a,goban13a,lodahl15a,goban15a,sipahigi16a}, namely, cold atoms in state dependent optical lattices~\cite{devega08a,navarretebenlloch11a}. The simplest scenario consists of a bosonic atom with two metastable states, $a$/$b$, trapped in a shallow/deep optical potential. In that situation, only the $a$-atoms tunnel to nearest neighbour at a rate, $J$, of the order of $10$ KHz in typical experiments~\cite{bloch08a}. Thus, the $a/b$ atoms play the role of the propagating bath modes/QEs respectively. Their coupling is obtained through either two-photon Raman transition or, in the case of Alkaline-Earth atoms, through direct optical coupling and can be $g\sim O(J)$, which allows one to tune $g$ and $\Delta$, from $0$  to several MHz, and thus investigate all the parameter regimes. Atomic motion effects can be suppressed by cooling the atom initially to the ground state and operating in the Lamb-Dicke regime.  The scattering losses induced by the trapping potential, finite lifetime of the optically metastable state in Alkaline-Earth atoms, or other imperfections lead to decoherence in both the QE/bath modes which can be as small as $\Gamma_\mathrm{loss} \sim $ Hz~\cite{daley08a,schreiber15a}, such that $g/\Gamma_\mathrm{loss}\sim 10^4$. The impact of losses on the observation of the phenomenology has been considered in Ref.~\cite{gonzaleztudela17b}. The possibility of addressing single sites~\cite{weitenberg11a,endres13a} and distinguishing the internal states, makes this setup ideal to observe our predictions. 

To sum up, we have explored the non-perturbative QEs dynamics emerging from their coupling to a two-dimensional bosonic bath with a square-like geometry when their energies lie in the middle of the band. For a single QE, we predict an exponential relaxation at short times, but with a timescale that escapes Fermi's Golden Rule description, which is followed by reversible and slow relaxation dynamics at longer times. Moreover, such phenomena are accompanied by strongly directional emission into the bath along two quasi-1D orthogonal directions, which lead to super/subradiant states when many QEs are coupled to the bath. For two QEs, the perturbative predictions are corrected due to the divergent density of states in the middle of the band, which forbids perfect super/subradiance. We characterize mathematically the phenomena and give an intuitive explanation in terms of interference. This understanding allows us to build perfect subradiant states with four QEs where the emission is trapped 
within them in spite of the 2D character of the bath.

Both the directionality and the divergent density of states responsible for the phenomena we describe, are associated to the saddle points in $\omega(\kk)$, and not its interplay with polarization as in 1D chiral systems \cite{lodahl16a}. Since those points are ubiquitous in 2D reservoirs, we conjecture that our findings will be relevant in more general situations beyond the simple model employed here~\cite{mekis99a}. Apart from the platforms mentioned, the predictions can also be tested in circuit QED~\cite{astafiev10a,hoi11a,vanloo13a,liu17a}.

\acknowledgements{\emph{Acknowledgements.} The work of AGT and JIC was funded by the EU project SIQS and by the DFG within the Cluster of Excellence NIM. AGT also acknowledges support from Intra-European Marie-Curie Fellowship NanoQuIS (625955). We also acknowledge discussions with T. Shi, Y. Wu, S.-P. Yu, J. Mu\~{n}iz, and H.J. Kimble.}

\bibliography{Sci,books}

\end{document}